\documentclass[aps,showkeys,showpacs,preprint]{revtex4-1}
\usepackage[dvips]{graphicx}

\begin{document}

\title{Cosmogenic Nuclei Production Rate on the Lunar Surface}
\author{ DONG Tie-Kuang $^{1,}$\footnote{E-mail address:tkdong@pmo.ac.cn }, YUN Su-Jun$^{2}$,
 MA Tao $^{1}$, CHANG Jin$^{1}$, DONG Wu-Dong$^{3}$,  ZHANG Xiao-Ping$^{3}$,
 LI Guo-Long$^{4}$, REN Zhong-Zhou $^{4,5}$}
\address{$^{1}$ Purple Mountain Observatory, CAS, 2 West Beijing Road, Nanjing 210008, China\\
$^{2}$ Nanjing XiaoZhuang University, 3601 Hongjing Road, Nanjing 211171, China\\
$^{3}$ Space Science Institute, Macau University of
Science and Technology, Avenida Wai Long, Taipa, Macau\\
$^{4}$Department of Physics, Nanjing University, Nanjing 210008, China\\
$^{5}$Center of Theoretical Nuclear Physics, National Laboratory
of Heavy-Ion Accelerator, Lanzhou 730000, China} \noindent

\begin{abstract}
A physical model of Geant4-based simulation of galactic cosmic ray
(GCR) particles interaction with the lunar surface matter has been
developed to investigate the production rate of cosmogenic nuclei.
In this model the GCRs, mainly very high energy protons and $\alpha$
particles, bombard the surface of the Moon and produce many
secondary particles such as protons and neutrons. The energies of
proton and neutron at different depths are recorded and saved into
ROOT files, and the analytical expressions for the differential
proton and neutron fluxes are obtained through the best-fit
procedure under the ROOT software. To test the validity of this
model, we calculate the production rates of long-lived nuclei
$^{10}$Be and $^{26}$Al in the Apollo 15 long drill core by
combining the above differential fluxes and the newly evaluated
spallation reaction cross sections. Numerical results show that the
theoretical production rates agree quite well with the measured
data. It means that this model works well. Therefore, it can be
expected that this model can be used to investigate the cosmogenic
nuclei in lunar samples returned by Chinese lunar exploration
program and can be extended to study other objects, such as the
meteorites and the Earth's atmosphere.
\end{abstract}



\maketitle

\section{Introduction}
Galactic Cosmic Rays (GCRs) are high energy particles that pass
through the interstellar space in our Milky Way galaxy. The most
abundant components of GCRs are protons and $\alpha$ particles.
These energetic particles can interact with the matters both in the
interstellar space and interplanetary space. The former can be used
to investigate the origin and evolution of GCR
itself\,\cite{strong2007}, and the latter provide us an useful tool to
investigate the cosmic ray exposure history of extraterrestrial
bodies such as the planets and the
meteorites\,\cite{reedy1983,ouyang1984,vogt1990,heusser1996,eugster2003}.
Among these bodies the Moon is an unique one in our solar system.
Due to the atmosphere free and old surface, the Moon has recorded
rich information about the solar system evolution and variations of
the galactic cosmic rays (GCRs) in the past several billion years.
GCRs can bombard directly the solid surface of the Moon. The
interactions of GCRs (primary particles) and the matters of the
lunar surface can produce many protons, neutrons and mesons
(secondary particles) and some nuclei that can not be synthesized by
thermonuclear reactions. These nuclei are called cosmogenic nuclei.
The materials returned by Apollo and Luna missions has provided good
opportunities to study the cosmogenic nuclei with very high
accuracy. It can be expected that the lunar samples returned by
Chinese lunar exploration program in the near future will promote
enormously the development of lunar and planetary science including
the investigation of cosmogenic nuclei. To provide reliable method
to interpret the measured data in the near future, it is time to
investigate this subject beforehand from theoretical points of view.

In principle, if both the concentration and production rate of a
stable cosmogenic nuclei are known, one can calculate the cosmic-ray
exposure (CRE) age of that material. The concentration of nuclei can
be measured accurately in laboratory by accelerator mass
spectrometer (AMS), but the calculation of the absolute production
rate is much more difficult in general speaking. Fortunately, there
is an important exception. If the CRE age is much longer than the
half-life of a given radioactive cosmogenic nucleus, its production
rate equals to the activity that can be measured directly (see
below). Therefore, most of the CRE ages are determined by using the
radioactive-stable nuclei pairs, for instance $^{10}$Be-$^{21}$Ne,
$^{26}$Al-$^{21}$Ne, $^{36}$Cl-$^{36}$Ar\,\cite{eugster2003}. The
CRE ages of lunar rocks on the rim of lunar craters are particularly
important to date the impact events that excavated these rocks to
GCR irradiation\,\cite{eugster2003}. These ages can be used to
calibrate the relative ages of lunar surface determined by the
morphology craters\,\cite{neukum2001}.

In this work, we develop a Monte Carlo model based on Geant4
software package to calculate the production rate of cosmogenic
nuclei. Before applying this model to extensive studies, it is
needed to test the validity of it. To this end, we choose the Apollo
15 long drill core as research object. It is because its chemical
composition, location, shielding depth of samples are known clearly.
In addition, Apollo 15 drill core is the least disturbed sample
among the three deep drill cores, Apollo 15, 16 and 17. The
production rates of long-lived nuclei $^{10}$Be and $^{26}$Al are
calculated and compared with experimental data. This paper is
oganized as follows: in section 2 we describe the theoretical model,
and in section 3 we discuss the numerical results, a summary and
outlook is given in section 4.

\section{The theoretical model}
The shape and size of meteoroids (i.e., the precursor of the
meteorites) are difficult to determine directly from that of
meteorites due to the unknown degree of ablation and fragmentation.
For simplicity, the meteoroids are usually regarded as a sphere with
radius $R$ and uniform chemical composition. Under this assumption,
the general formula of the production rate of cosmogenic nuclei
$i$ at the depth $d$ under the surface of a meteoroid is shown as
follows\,\cite{arnold1961,reedy1972}:
\begin{eqnarray}
P_{i}=\sum _{k} \sum_{j} N_{j} \int  J_{k}(E,R,d)\sigma _{j,k}(E)
dE,
\end{eqnarray}
where $N_{j}$ is the concentration of element $j$ in the meteoroid,
$J_{k}(E,R,d)$ is the differential flux of particle $k$ (in this work we only consider
protons and neutrons), $\sigma_{j,k}(E)$ is the excitation function
of the involved nuclear reaction for the production of nuclide $i$
from element $j$ induced by particle $k$ at incident energy $E$.

Before the meteoroids impacting the Earth's atmosphere, they are
irradiated by galactic cosmic rays. In this work, we take the most
abundant proton and $\alpha$ particles in GCRs as primary particles,
which bombard the surface of meteoroids isotropically. The fluxes
of proton and $\alpha$ particle are taken as
follows\,\cite{masarik2009}:
\begin{eqnarray}
J_{p}(E,M)=C_{p}\frac{E(E+2m_{p}c^{2})} {(E+M)(E+M+2m_{p}c^{2})}
(E+M+x_{p})^{-\gamma_{p}},
\end{eqnarray}
\begin{eqnarray}
J_{\alpha}(E,M)=C_{\alpha}\frac{E(E+2m_{p}c^{2})}{(E+M)(E+M+2m_{p}c^{2})}
(E+M+x_{\alpha})^{-\gamma_{\alpha}}.
\end{eqnarray}
In these two formulas $E$ denotes the energy per nucleon in MeV, $M$
is the solar modulation parameter in MV, and $m_{p}c^2=938$ MeV is the rest mass of proton. In
Eq.(2) $C_{p}=1.244 \times 10^{6}$(cm$^{2}$ s MeV)$^{-1}$, $x_{p}=
780~\text{exp}(-2.5\times10^{-4} (E+M))$, $\gamma_{p}=2.65$. In
Eq.(3) $C_{\alpha}=2.23 \times 10^{5}$(cm$^{2}$ s MeV)$^{-1}$,
$x_{\alpha}= 660~\text{exp}(-1.4\times10^{-4} (E+M))$,
$\gamma_{\alpha}=2.77$. These formulas are implemented into the
Geant4 code to sample the primary protons and $\alpha$ particles.

When the high energy protons and $\alpha$ particles bombard the
meteoroid, they will interact with the atoms and nuclei therein. The
dominant processes are ionizing energy loss and spallation
reactions. By ionizing energy loss the energy of primary particles
will decrease continuously. And by the spallation reactions many
secondary particles, such as protons, neutrons and mesons, are
produced. The secondary particles, especially neutrons will induce
lower-energy ($E<100$ MeV) reactions, and the very low-energy neutrons will
account for neutron capture reactions.

As shown in Eq.(1), in order to calculate the production rate one
must search for a method to obtain the differential fluxes of proton
and neutron in the irradiated body. The Monte Carlo simulation is a
good choice for these calculations. Among various Monte Carlo based
software, the Geant4\,\cite{ago2003} has been used more and more
widely in nuclear physics and space
science\,\cite{song2009,zhang2009}. In our recent
work\,\cite{dong2011}, the neutron fluxes escaped from the lunar
surface has been investigated using the Geant4 software. In the
present work we will also use this software to calculate the proton and neutron
fluxes beneath the meteoroid surface. In order to obtain the
differential fluxes of active particles, the energies of primary and
secondary proton and neutron at different depths are recorded and
saved as ROOT files. Then we obtain the numerical differential
fluxes of different particles by analyzing the ROOT file. However,
due to the statistical nature of the Monte Carlo method, there are
statistical fluctuations in the fluxes. To calculate the production
rate of cosmogenic nuclei by Eq.(1), it is necessary to obtain the
analytical expressions of proton and neutron fluxes by the best-fit procedure.
Once the analytical expressions for fluxes are
obtained, one can calculate the production rate of interesting
nuclei by combining these fluxes formulas with the nuclear reaction
cross sections. The cross sections of proton induced reactions can
be taken from experimental data. However, the cross sections for
high energy neutron induced reactions are obtained in indirect ways,
for instance the thick target irradiation experiment. It is because
the neutron is a neutral particle that can not be accelerated to any
energy. In this paper we will use the evaluated proton cross
sections by Nishiizumi et al.\,\cite{nish2009} and neutron cross
sections by Reedy\,\cite{reedy2013}.

Up to now, the physical model to calculate the production rate of
cosmogenic nuclei has been established. After apply this model to
extensive studies, it is needed to test the validation of this
model. To this end, we choose the lunar drill core as research
object. It is because its chemical composition, location, shielding
depth of samples are known clearly. In addition, Apollo 15 drill
core is the least disturbed sample among the three deep drill cores,
Apollo 15, 16 and 17\,\cite{nish1984a,nish1984b}. In this case the
radius of Moon can be regarded as infinite, and the surface can be
seen as a plane. In simulations we establish a sufficient large box
filled with the average chemical composition of Apollo 15 drill
core\,\cite{gold1977}, the GCRs incident on the upper surface of it.
In this work, the box is taken as 100 m long $\times$ 100 width
$\times$ 20 m thickness. This box is filled with uniform lunar
soils. The chemical composition is: O (43.0\% wt), Si (22.1\% wt),
Al (7.67\% wt), Ca (7.52 \% wt), Mg (5.89 \% wt), Fe (11.57 \% wt), Ti
(1.12 \% wt), Na (0.332 \% wt), and K (0.21 \% wt). Here, \% wt denotes
the weight percentage. The GCRs, sampled according to Eqs.(2) and (3), incident
on the upper surface of this box isotropically. Considering both the
statistical accuracy and time cost, 5 million particles are sampled.

Once the production rate is calculated one can derive the cosmic ray
exposure (CRE) age by using the relations as follows: for a stable
cosmogenic nuclide
\begin{eqnarray}
S=P_{s}  T ,
\end{eqnarray}
and for a radioactive cosmogenic nuclide with the decay constant $\lambda$
\begin{eqnarray}
R=P_{R} \lambda ^{-1} (1-e^{-\lambda T}),
\end{eqnarray}
where $T$ is the CRE age and  $S$ ($R$) and $P_{s}$ ($P_{R}$) are the concentration and production
rate of stable (radioactive) nuclide. The basic unit of the cosmogenic nuclei concentration
is the number of atoms per gram sample (atoms/(g sample)). In literatures, the
concentrations of radioactive nuclei are often reported as radioactive activity, in unit of
disintegrations per minute per kilogram (dpm kg$^{-1}$).
The relationship between the concentration and the activity is
\begin{eqnarray}
A=\lambda R =P_{R}(1-e^{-\lambda T}).
\end{eqnarray}
From this equation, one can see that if the CRE age is much longer
than the half-life of this nuclide, the radioactive equilibrium will
be reached and the activity is approximately equal to the production
rate. For this case the measured quantity can be compared directly
with the calculated ones. In this paper, we will choose the nuclei
whose half-lives are much longer than the time by which the lunar
core are returned from the Moon and much shorter than the CRE age.
$^{10}$Be ($t_{1/2}=1.51 \times 10^{6} yr$), $^{26}$Al
($t_{1/2}=7.17 \times 10^{5} yr$) are good candidates.

\section{Numerical results and discussions}
According to the theoretical model described in Sec.\,2, we write
the Geant4 code to simulate the transportation and interaction of
GCRs in the lunar surface. In simulations, the neutrons and protons
at different depths are recorded and saved as ROOT files, from which
we obtain the differential fluxes for many energy bins. Then the
differential fluxes for neutron and proton at different energy ranges
are fitted, respectively. After careful analysis we found that the
proton and neutron fluxes can be expressed quite well by the
following formulas:
\begin{eqnarray}
J_{p}(E,d)= \left \{ \begin{array}{ll}
p_{0} (E+p_{1})^{p_{2}} E/(E+p_{3}), ~~~ (10 \leq E \leq 500)  \\
p_{0} (E+p_{1})^{p_{2}}, ~~~~~~~~~~~~~~~~~ (500 \leq E \leq 10^{4})
 \end{array} \right .
\end{eqnarray}
\begin{eqnarray}
J_{n}(E,d)=  \left \{ \begin{array}{ll}
p_{0} e^{-E/p_{1}}, ~~~~~~~~~~~~~~~~~~~~~~  (10 \leq E \leq 200)  \\
p_{0} e^{-E/p_{1}}/(E+p_{2}), ~~~~~~~~~  (200 \leq E \leq 10^{4})
 \end{array} \right .
\end{eqnarray}

In these formulas the energy $E$ is in MeV, and the $p$'s are depth
and chemical dependent parameters. These formulas are much simpler than that used by
Arnold\,\cite{arnold1961} and Reedy\,\cite{reedy1972}. The parameters are determined by
the best-fit procedure using the ROOT software. Considering the
limit of space we will not give the numerical results of theses
parameters. The proton and neutron fluxes calculated using Eqs.(7)
and (8) at depths $d=100,200,300,400 $ g/cm$^{2}$ are shown in
Fig.\,1. Here the depth $d$ is the geometrical depth times the density, the unit is g/cm$^{2}$.
 In this figure the fluxes are not normalized to
particles/(cm$^{2}$ s MeV) since we are interested to the relative
magnitude of proton and neutron fluxes. From this figure, one can
see that at high energy range ( $E
>200$ MeV) the proton flux is larger than that of neutron, but at
low energy range ($E<100$ MeV) the neutron flux is larger than that of proton.
It is because the protons beneath the lunar surface include the
primary and secondary ones, but the neutrons are purely secondary
particles. The secondary protons will suffer ionizing loss and stop
quickly.
\begin{figure}[htb]
\centering
\includegraphics[width=10cm]{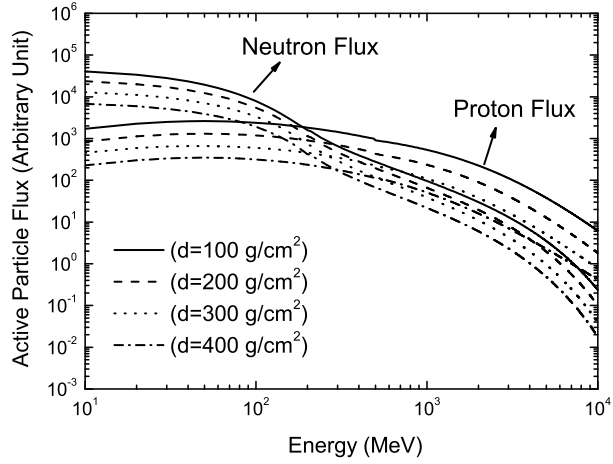}
\caption{  The proton and neutron fluxes at depths
$d=100,200,300,400 $ g/cm$^{2}$.}
\end{figure}
As for the cross sections for proton and neutron induced reactions,
the most newly evaluated data are used. The proton cross sections are taken
from\,\cite{nish2009}, and the neutron cross sections are taken
from\cite{reedy2013}.  The numerical data for the cross sections involved
are shown in Figs.2-4.

\begin{figure}[htb]
\centering
\includegraphics[width=10cm]{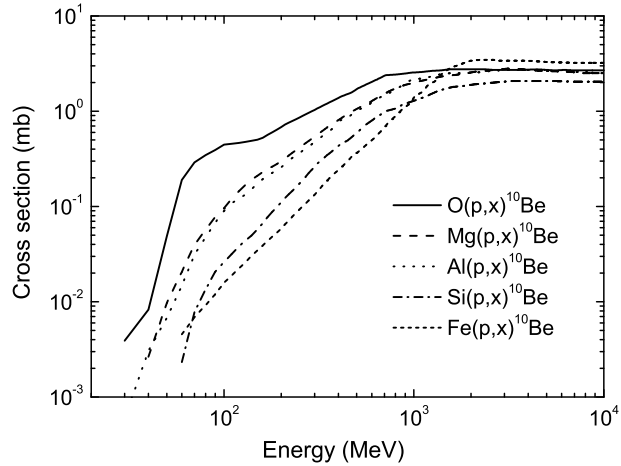}
\caption{ The excitation function for proton induced
reactions to produce $^{10}$Be on target elements O, Mg, Al, Si, Fe.
These data are taken from\,\cite{nish2009}.}
\end{figure}
\begin{figure}[htb]
\centering
\includegraphics[width=10cm]{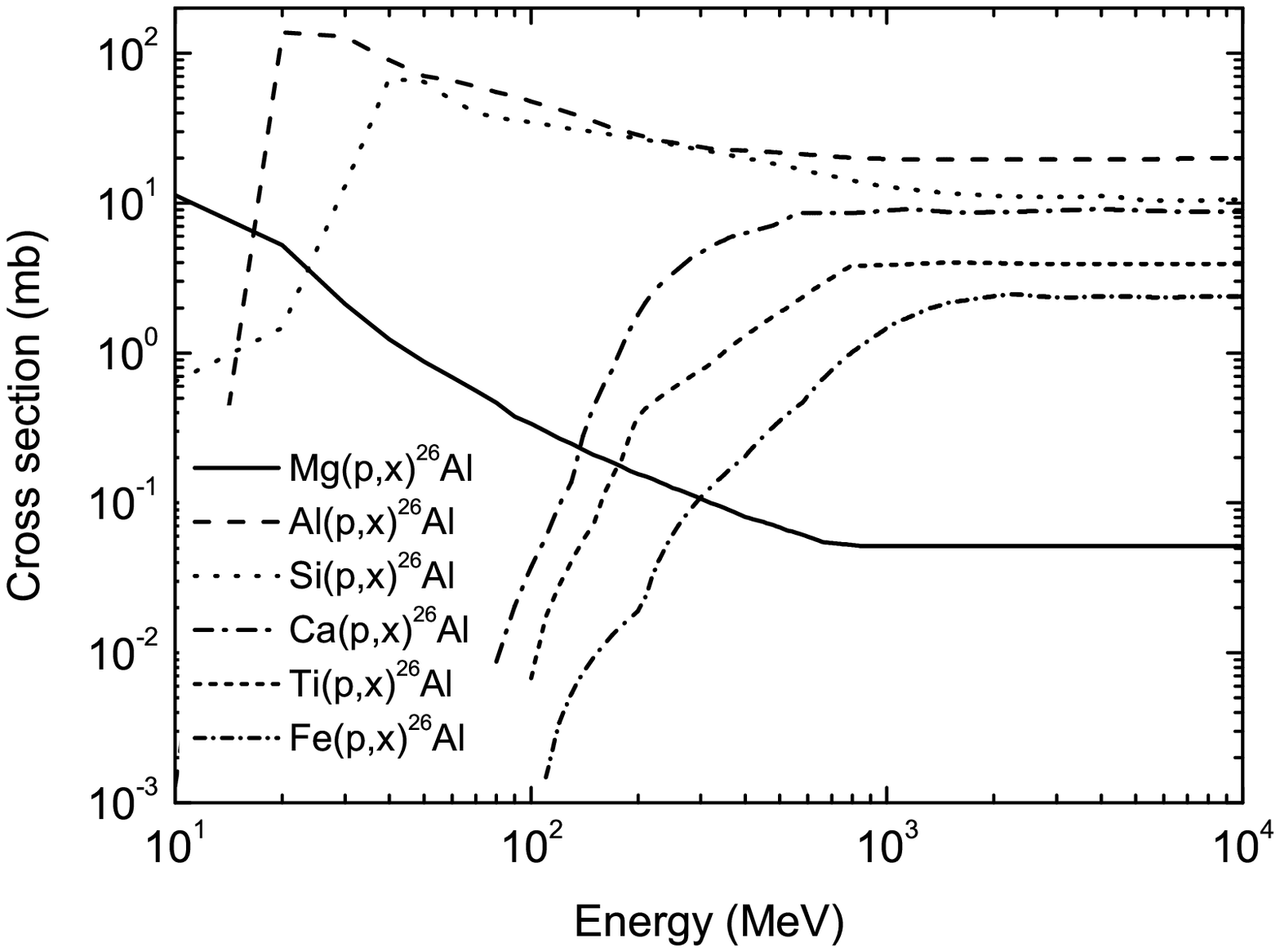}
\caption{\label{fig3} The excitation function for proton induced
reactions to produce $^{26}$Al on target elements Mg, Al, Si, Ca,
Ti, Fe. These data are taken from\,\cite{nish2009}.}
\end{figure}
\begin{figure}[htb]
\centering
\includegraphics[width=10cm]{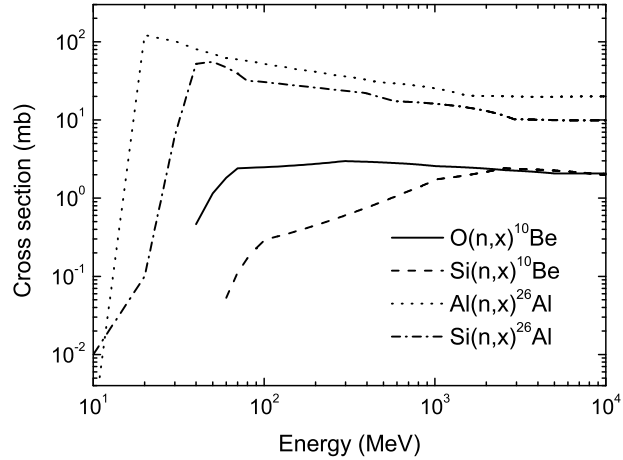}
\caption{ The excitation function for neutron
induced reactions to produce $^{10}$Be on target elements O, Si, and
$^{26}$Al on elements Al, Si. These data are taken
from\,\cite{reedy2013}.}
\end{figure}

From these figures one can see clearly that the excitation functions
for these reactions are not smooth. It is because the spallation
reaction is a very complex process, the spallation
cross sections are very difficult to calculate. Therefore, these
functions are obtained by interpolating the experimental data.
Unfortunately, there are only a few data available, even for neutron
induced reactions.
\begin{figure}[htb]
\centering
\includegraphics[width=10cm]{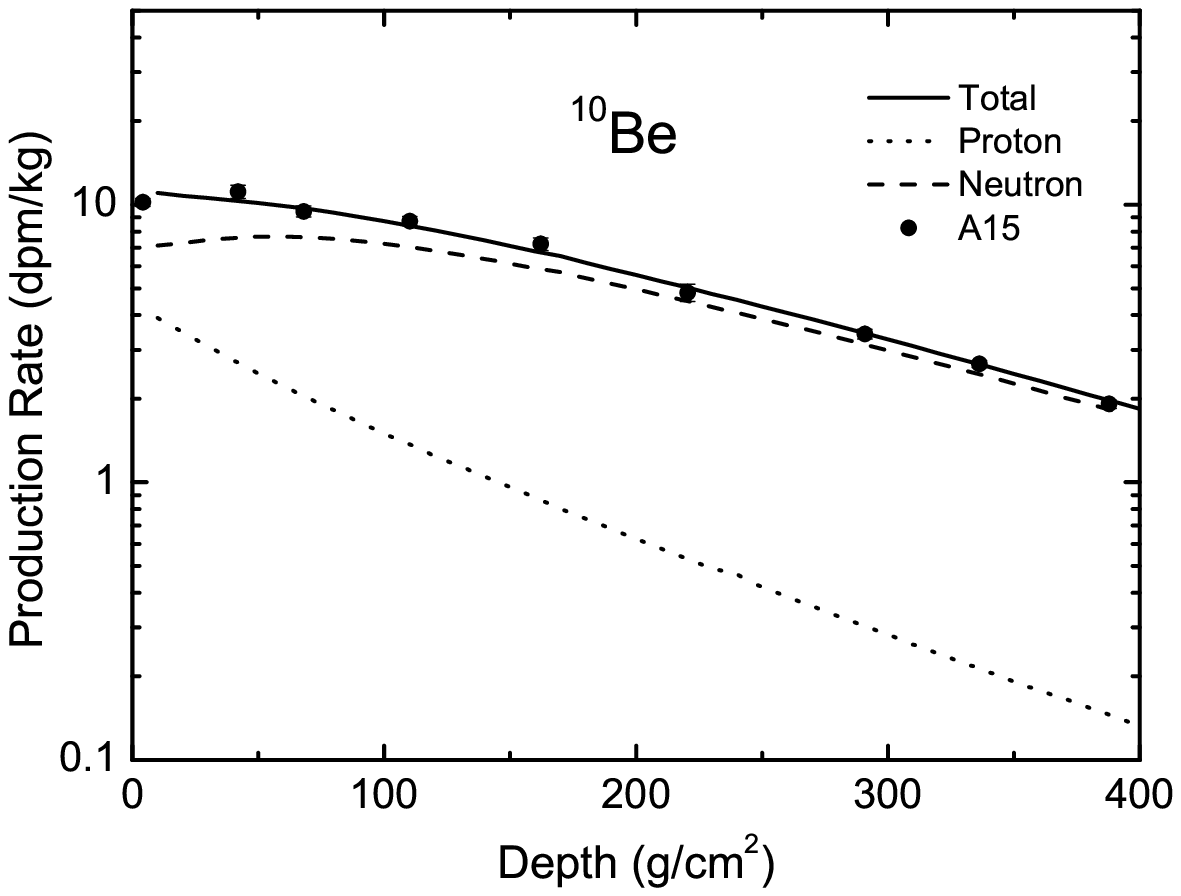}
\caption{ The theoretical and experimental production
rate of $^{10}$Be in Apollo 15 drill core. The experimental data are
taken from \cite{nish1984a}. }
\end{figure}
\begin{figure}[htb]
\centering
\includegraphics[width=10cm]{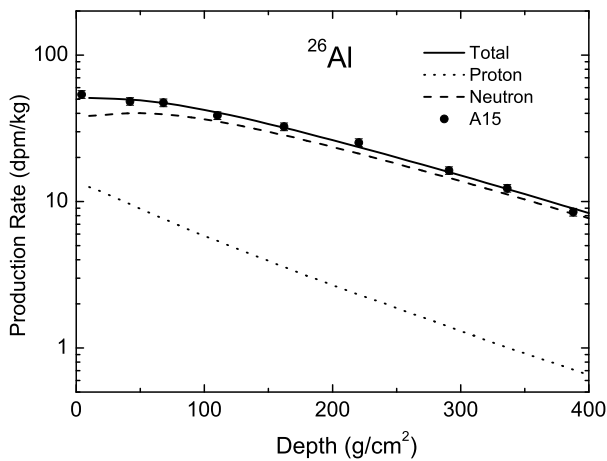}
\caption{ The theoretical and experimental production
rate of $^{26}$Al in Apollo 15 drill core. The experimental data are
taken from \cite{nish1984b}.  }
\end{figure}

Finally, the production rates of cosmogenic nuclei $^{10}$Be and $^{26}$Al are calculated by
integrating the depth-dependent fluxes of active particles with the
excitation functions of nuclear reactions involved. By using this
method, the production rate of $^{10}$Be and $^{26}$Al in Apollo 15
drill core are calculated. In calculations, the production rates
result from proton induced reactions and neutron induced reactions
are considered.  As for proton induced reactions the target elements
we considered are O, Mg, Al, Si, Fe for $^{10}$Be, Al, Si, Ca, Ti,
Fe for $^{26}$Al. As for neutron induced reactions the cross
sections are very scattered. The target elements we considered are O
and Si for $^{10}$Be, Al and Si for $^{26}$Al. In order to
compare with experimental data,
we introduce two normalization factors for the
production rates for neutron and proton induced reactions:
\begin{eqnarray}
P_{R}=C_{p} P_{R,p}+C_{n} P_{R,n},
\end{eqnarray}
where $P_{R}$ is the total production rate of radioactive nuclei,
$P_{R,p}$ and $P_{R,n}$ are the contributions due to proton and
neutron induced reactions, respectively, calculated using the
theoretical model shown in Sec.\,2. The main reasons of introducing
these factors are as follows: (1) the excitation functions are
obtained by interpolating a small number of cross section data,
particularly the neutron cross section data are obtained by indirect
method; (2) the secondary proton and neutron yields calculated by
Geant4 depend on the theoretical model of spallation reactions; (3)
the spectrum of primary cosmic rays are assumed to be the current
mean value averaged during the solar activity cycle.  That is to say
the uncertainties on the primary GCR fluxes, secondary particle
fluxes and excitation functions in the theoretical model are
normalized by two parameters $P_{R,p}$ and $P_{R,n}$.  These two
parameters can be obtained by the best-fit procedure. The
theoretical and experimental results are shown in Figs.5-6. In these
figures the dotted line and dashed line represent the contributions
from proton and neutron induced reactions, respectively, and the
solid lines denote the total production rate. The experimental data
(denoted by circles) for $^{10}$Be are taken from \cite{nish1984a},
and the data for $^{26}$Al are taken from\cite{nish1984b}. From
these two figures one can see that the theoretical production rate
of $^{10}$Be and $^{26}$Al agree well with the measured data. In
addition, the production rate from neutron induced reactions are
larger than that of proton induced reactions. In other words, the
neutron reactions dominate the production of cosmogenic nuclei,
hence the high-energy neutron reaction cross sections play crucial
roles.

\section{Summary and Outlook}
In summary, a physical model based on the Monte Carlo method is
proposed to calculate the production rate of cosmogenic nuclei. A
Geant4 code has been developed to simulate the transport and
interactions of the primary and secondary particles in the
extra-terrestrial body. As a test of this model we calculate the
production rate of long-lived nuclei $^{10}$Be, $^{26}$Al in the
Apollo 15 drill core. The information of neutrons and protons at
different depths are recorded and saved as ROOT file. Then the
neutron and proton differential fluxes are obtained by using ROOT
software. By combining the differential fluxes and the newly
evaluated spallation cross sections the production rates of
$^{10}$Be, $^{26}$Al are calculated and compared with experimental
data. The results show that the theoretical production rates agree
quite well with the measured data for the Apollo 15 long drill core.
It means that our model is suitable to investigate the production
rate of cosmogenic nuclei in extra-terrestrial body. It is
reasonable to believe that this model can be applied to  the
investigation of production rate of cosmogenic nuclei in  meteorites
and Earth's atmosphere. Furthermore, according to the Chinese lunar
exploration program, lunar samples will be returned 10 years later
or so. When lunar samples are obtained the cosmic ray exposure
history of the landing sites is an important subject of research.
For instance, the cosmic ray exposure age of the crater rims can be
used to date the impact events and to calibrate the geological age
of that region defined by the counting rate of impact craters. To
reach this goal, it is needed to improve this model by using better
physical model of high energy nuclear reactions and more accurate
cross sections of proton and neutron induced reactions. As for the
former, we can expect that with the improvement of the nuclear
reaction model, for instance Li$\grave{e}$ge intranuclear cascade
(INCL) model\,\cite{boudard2012}, the proton and neutron spectrum
can be calculated more and more accurate. In the next, we will study
the influences of different nuclear reaction models on the proton
and neutron fluxes. As for the latter, the neutron cross sections
measured by using the quasi-monoenergetic neuron beams produced by
$^{7}$Li(p,n)$^{7}$Be reaction within the European HINDAS (High- and
INtermediate-energy Data for Accelerator-driven Systems)
project\,\cite{koning2002} will give us unprecedented opportunity to
study the production rate of cosmogenic nuclei.

\begin{center}
{\large Acknowledgments }
\end{center}
This work is supported by the National Natural Science Foundation of China(Grant
No.11303107, 11105079), by the Science and Technology of Development
Fund of Macau (Grant No.068/2011/A).

\clearpage

\end{document}